\documentclass[%
 reprint,
amsmath,amssymb,
aps,prd
]{revtex4-2}

\usepackage{graphicx}
\usepackage{bm}
\usepackage{booktabs}
\usepackage{amsmath}
\usepackage{subfigure}
\usepackage{caption}
\usepackage[hidelinks]{hyperref}
\hypersetup{
    colorlinks=true,
    linkcolor=blue,
    filecolor=gray,      
    urlcolor=blue,
    citecolor=blue,
}

\begin{document}


\title{Gravitational wave signatures and detectability of the mass transfer effect in compact binaries }

\author{Zi-Han Zhang$^{1,2,3,4}$}
 \email{zhangzihan242@mails.ucas.ac.cn}

\author{Tan Liu$^{1,2}$}
 \email{lewton@mail.ustc.edu.cn}

\author{Shenghua Yu$^{5}$}
 \email{shenghuayu@bao.ac.cn}

\author{Zong-Kuan Guo$^{1,2,6}$}
\email{guozk@itp.ac.cn}

\affiliation{$^{1}$School of Fundamental Physics and Mathematical Sciences, Hangzhou Institute for Advanced Study, University of Chinese Academy of Sciences, Hangzhou 310024, China}
\affiliation{$^{2}$School of Physical Sciences, University of Chinese Academy of Sciences, Beijing 100049, China }
\affiliation{$^{3}$International Centre for Theoretical Physics Asia-Pacific, University of Chinese Academy of Sciences, 100190 Beijing, China}
\affiliation{$^{4}$Taiji Laboratory for Gravitational Wave Universe, University of Chinese Academy of Sciences, 100049 Beijing, China}
\affiliation{$^{5}$CAS Key Laboratory of FAST, National Astronomical Observatories, Chinese Academy of Sciences, 20A Datun Road, Beijing 100101, China}
\address{$^{6}$CAS Key Laboratory of Theoretical Physics, Institute of Theoretical Physics, Chinese Academy of Sciences, Beijing 100190, China }

\date{\today}

 \begin{abstract}
 The mass transfer process is prevalent during the inspiral phase of compact binary systems. Our study focuses on systems comprising low-mass white dwarfs, particularly in neutron star-white dwarf binaries and double white dwarf binaries, where a stable mass transfer process occurs at low frequencies. By analyzing the evolution of gravitational wave frequencies in the presence of mass transfer within quasi-circular orbits, we derive an analytical expression for the time-dependent frequency across different frequency bands and the waveforms emitted by compact binaries. Considering gravitational waves emitted by compact binaries in the $1\thicksim10$ mHz band, based on the Fisher analysis, we find that the mass transfer rate can be measured as accurately as $10^{-7} M_\odot/\text{year}$ by space-based gravitational-wave detectors with a signal-to-noise ratio of the order of $10^3$. Including the mass transfer effect in the waveforms provides a new possibility to measure the individual masses of double white dwarf binaries. The relative error of measured white dwarf masses can be down to the order of 0.01.
 \end{abstract}
 
 \maketitle
 
\section{Introduction}
Gravitational waves (GWs) emitted by inspiralling compact binaries have been detected by LIGO/Virgo. The space-based GW detectors LISA, Taiji, and Tianqin \cite{PhysRevLett.100.041102, PhysRevD.102.063021,Kang_2021} are all sensitive to frequencies around 10 mHz \cite{babak2021lisasensitivitysnrcalculations}, a band where GWs from these inspiralling binaries exhibit the highest SNR \cite{Babak_2006,PhysRevD.106.102004,PhysRevD.104.024023}. Compact binary systems are abundant in the Milky Way, particularly double neutron star (NS-NS), neutron star-white dwarf (NS-WD), and double white dwarf (WD-WD) systems, with an estimated event rate of approximately $10^4$ per year\cite{PhysRevLett.128.041101}.

The accretion process is a common phenomenon in binary systems containing white dwarfs (WDs). Due to the density differences between the binary stars, matter from the lower-density star is transferred to the higher-density star, a process known as mass transfer (MT) \cite{PhysRevD.109.123013,Fernandez_2013,Chen_2023}. These MT processes can delay or even interfere with the merger of binary systems \cite{Accretion_power}. Under the assumption of total mass conservation, $ M = m_1 + m_2 $, this study incorporates the effect of MT into GWs emitted by the compact binaries \cite{10.1093/mnras/stab626,1988ApJ...324..851M}. During the inspiral phase, the MT rate is relatively low, and the accretion rate must remain below the Eddington limit to maintain mass conservation. In the case of quasi-circular orbits, the stable MT rate can be approximated as constant over a short observation period (approximately 10 years). This assumption simplifies the analytical calculations presented in this paper.

MT influences the mass distribution within binary star systems, thereby affecting both the orbital dynamics and the resulting gravitational waveforms \cite{1992ApJ...394..586J}. Its influence on the time derivative of the GW frequency can be divided into direct and indirect parts. The direct influence corresponds to the term in Eq.\eqref{GWf} that is proportional to the MT rate. The indirect part corresponds to the gravitational radiation back-reaction in Eq.\eqref{GWf}, since this term depends on the masses of the stars. However, the direct influence of MT is ignored in \cite{2024arXiv240216612L} and the indirect influence is ignored in \cite{10.1093/mnras/stad2358}. 

In our work, we utilize perturbation theory to derive an analytical expression for the time evolution of the GW frequency. In Sec.\ref{Sec2}, we employ the time derivative of the GW frequency for quasi-circular orbits with MT and discuss the MT rate and its applicable range. There is a critical frequency such that the gravitational radiation back-reaction the MT effect balance, and the time derivative of the GW frequency is zero.  When the frequency is below the critical frequency, MT has a more pronounced effect than the gravitational radiation back-reaction, causing the binaries to drift apart; when the frequency is higher than the critical frequency, the stronger gravitational radiation back-reaction drives them closer together. Consequently, the frequency may increase or decrease depending on the specific frequency band. In Sec.\ref{Sec3}, we get the analytical frequency evolution function in the low-frequency case and compare it with numerical results obtained when the MT rate is constant. We also get the Newtonian quadrupole order gravitational waveforms.  Over a 2-year observation period, the impact of MT on the GW phase is illustrated in Fig.\ref{error}.
In Sec.\ref{Sec4}, using the Newtonian quadrupole order waveforms emitted by the compact binaries, we employ the Fisher analysis to compute the projected parameter uncertainties of the GW detection with Taiji and LISA.  We conclude and point out future directions in Sec.\ref{sec5}.


\section{GW Frequency with MT}\label{Sec2}
\begin{figure}[ht!]
    \centering
    {\includegraphics[width=0.48\textwidth]{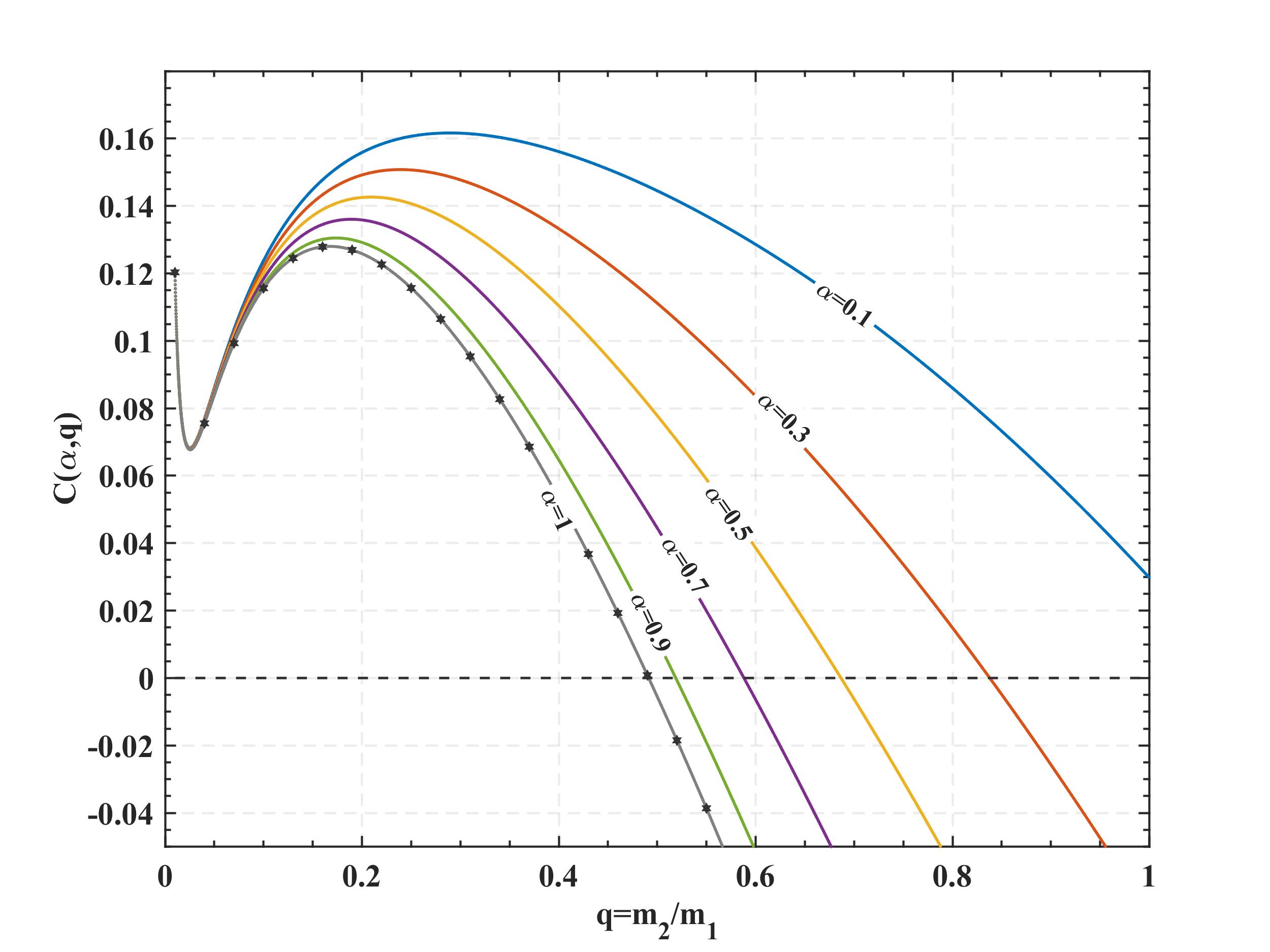}}
    \caption{Function $\mathcal{C}$ of $q$ with different values of $\alpha$. We mainly consider the part $\mathcal{C}>0$ while mass ratio $q<0.5$, $\alpha=1$, and $\beta=1$.}
    \label{C_q}
\end{figure}

Considering a binary system with a primary star of mass $m_1$  and a companion star of mass $m_2$ in a quasi-circular inspiral orbit with frequency $f/2$, we define the mass ratio as $0 < q = m_2/m_1 < 1$ and the total mass as $M = m_1 + m_2$. The MT process and GW radiation can influence the orbital energy and angular momentum of the system, and the time derivative of the GW frequency is given by \cite{10.1093/mnras/stab626}
\begin{equation}
\frac{\dot{f}}{f}=\frac{96}{5}\pi^{8/3}c^{-5}G^{5/3}m_1m_2M^{-1/3}f^{8/3}+3\mathcal{C}\frac{\dot{m}_2}{m_2}.\label{GWf}
\end{equation}
Here $G$ is the gravitational constant and $c$ is the speed of light. The first term on the right side is due to the gravitational radiation and the second term represents the contribution due to MT. The MT effect indirectly influences the first term by changing $m_1$ and $m_2$. Astronomical fitting results for $\mathcal{C}$ indicate that the reduction in angular momentum is primarily due to the formation of an accretion disk and the total mass loss \cite{10.1093/mnras/stw1410}. The change in stellar masses is described by $\dot{m}_1 = -\alpha\dot{m}_2>0$, where $\alpha$ represents the MT efficiency \cite{1994A&A...288..475P}. Using $\beta$ to quantify the angular momentum loss due to mass ejection, $\mathcal{C}$ can be expressed as
\begin{equation}
    \mathcal{C}=1-\alpha q-\left(\beta+\frac{1}{2}\right)\left(1-\alpha\right)\frac{q}{1+q}-\alpha\sqrt{\left(1+q\right)r_h}.
\end{equation}
The relationship between  $\mathcal{C}, q$, and $\alpha$ is illustrated in Fig.\ref{C_q}. When $\mathcal{C} > 0$, MT causes the binary to widen its orbit, whereas when $\mathcal{C} < 0$, MT accelerates the binary's inspiral. In this study, we focus on the case where $q<0.4$ and $\mathcal{C} > 0$ to explore the equilibrium between the effects of MT and GW radiation. It can be seen that when the frequency is low enough, the MT effect dominates and the time derivative of the GW frequency can be negative.

In a mass-conservative binary system, where the total mass $M$ remains constant and $\dot{m}_1 = -\dot{m}_2$, the coefficient of angular momentum for the accretion disk is $r_h = 0.0883 - 0.04858\ln q + 0.11489\ln^2 q + 0.020475\ln^3 q$  \cite{1988ApJ...332..193V}, allowing  $\mathcal{C}$  to be expressed as a function of $q$. The variation of the mass ratio $q$ over one-year evolution is of the order $10^{-5}$ at the Eddington accretion limit. Therefore, $q$ can be treated as constant in the analytical approach. We verify this approximation by numerical calculations in Sec.\ref{ana_num}.

\subsection{Critical frequency}
For the star with mass $m_1$ and radius $R_*$, when the star absorbs mass at its surface, gravitational potential energy is released, this energy is eventually released as electromagnetic radiation such as X-rays  \cite{Accretion_power,2023pbse.book.....T}. When the electromagnetic radiation pressure is equal to the gravitational potential, this corresponds to a limit luminosity called the Eddington limit.

For the stable accretion, there is a stable accretion rate $\dot{m}_1$ at the Eddington limit. If the surface of the star absorbs all the mass, the accretion luminosity is
\begin{equation}
L_{acc}=Gm_1\dot{m}_1/R_{*}.
\end{equation}
Based on this equation, we can get the Eddington mass accretion rate   \cite{Accretion_power,2023pbse.book.....T} 
\begin{equation}
    -\chi=\dot{m}_1=4\pi GMcR_*/\sigma_T=2.2\times10^{-9} M_\odot/\text{year} \left(\frac{R_*}{\text{km}}\right),
\end{equation}
where $\sigma_T=6.7\times10^{-25} \text{cm}^2$ is the Thomson cross section. We mainly consider the radius $R_*=10\thicksim10^4$ km. We can use the radius-mass relation \cite{1972ApJ175417N} to turn the MT rate into a function of the mass $m_1$. In this situation, the accretion rate of typical neutron star $(R_*\thicksim 10~ \text{km})$ and white dwarf $(R_*\thicksim 10^4 ~\text{km})$ are $10^{-8} M_\odot/\text{year}$ and $10^{-5} M_\odot/\text{year}$, respectively.

Then we can solve  Eq.(\ref{GWf}) to obtain the time evolution of the frequency $f$.  Following previous studies of the GW waveforms with the MT effect \cite{2024arXiv240216612L,10.1093/mnras/stad2358}, we  set the accretion rate $\dot{m}_2\equiv\chi <0$ as a constant, thus the masses in the  binary system at any time are
\begin{equation}
    m_1=m_{1,0}-\chi t,\quad m_2=m_{2,0}+\chi t,\quad M=m_1+m_2,\label{mtrate}
\end{equation}
where $m_{1,0},\;m_{2,0}$ are the initial masses of the binary, and in the normal situation when $m_{1,0}>m_{2,0}$, the mass is transferred from $m_{2,0}$ to $m_{1,0}$.

We can rewrite the Eq.(\ref{GWf}) as 
\begin{equation}
    \frac{\dot{f}}{f}=\mathcal{F}\left(m_{1,0}-\chi t\right)\left(m_{2,0}+\chi t\right)f^{8/3}+3\mathcal{C}\frac{\chi}{m_{2,0}+\chi t},\label{ff}
\end{equation}
where the constant $\mathcal{F}=96/5\pi^{8/3}c^{-5}G^{5/3}M^{-1/3}$. In the low-frequency case, the MT effect dominates, while in the high-frequency case, GW radiation has a greater influence. Therefore, we can use perturbation methods to obtain low and high frequencies analytical solutions. Setting $\dot{f}=0$ in Eq.\eqref{GWf} yields the critical frequency
\begin{equation}\label{fc}
\begin{split}
      f_c=&11.5\text{mHz}\times\left(\frac{\chi}{-2.2\times 10^{-6} M_\odot/\text{year}}\right)^{3/8}\left(\frac{\mathcal{C}}{0.12}\right)^{3/8}\\
     &\times\left(\frac{m_{1}}{2M_\odot}\right)^{-3/8}\left(\frac{m_{2}}{0.2M_\odot}\right)^{-3/4}\left(\frac{M}{2.2M_\odot}\right)^{1/8},    
\end{split}
\end{equation}
and $f_c$ is the function of binary's masses and the MT rate in the stable MT process.

\subsection{Frequency in the Roche limit}
MT occurs when the distance between the binary stars is less than the Roche limit or when the radius of the companion star exceeds the Roche radius \cite{2023pbse.book.....T}
\begin{equation}
    R_{RLOF}=\frac{0.49q^{2/3}}{0.6q^{2/3}+\ln(1+q^{1/3})}a,
\end{equation}
where $a$ represents the semi-major axis of the binary system. For a quasi-circular orbit, the relationship between the GW frequency $f$ and the semi-major axis $a$ can be derived from Kepler's laws: $a = (GMf^{-2}/\pi^2)^{1/3}$. Using this relation, for the GW frequency $f = 10$ mHz and a primary star of mass $1\sim 2 M_\odot$, the WD is estimated to have a mass around $0.2 M_\odot$ such that its radius $1.46 \times 10^4$ km is equal to the Roche radius. Under these conditions, if the mass of the WD significantly exceeds $0.2 M_\odot$, the MT process would not occur. See also Table 1 in \cite{10.1093/mnras/stab626}.

\begin{figure}[ht!]
    \centering
    {\includegraphics[width=0.48\textwidth]{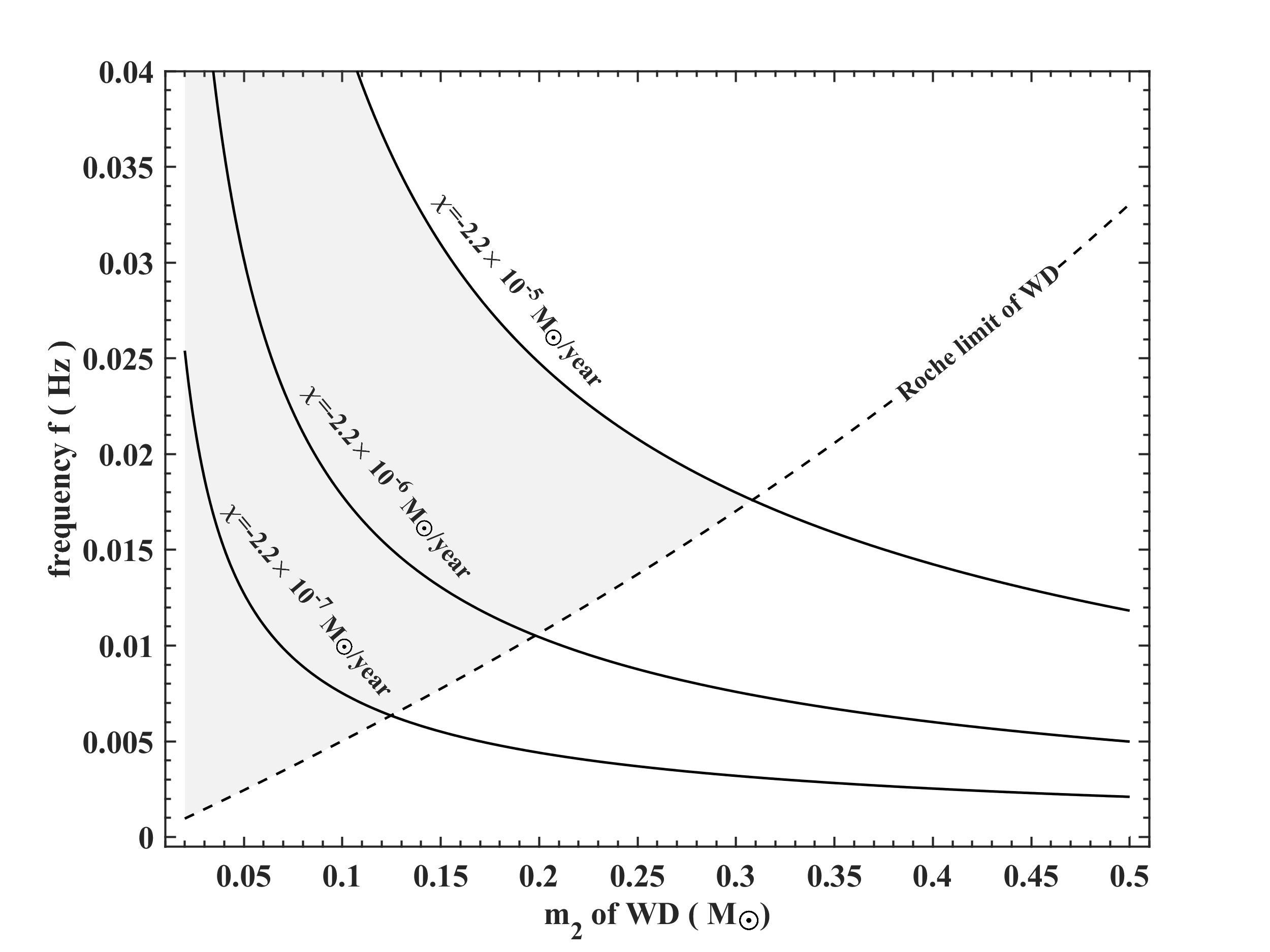}}    
    \caption{Range of the GW frequency such that the low-frequency approximation is applicable. The GW frequency should be lower than  $f_c$ and higher than $f_{RLOF}$. The dotted line shows the frequency $f_{RLOF}$ of Roche limit changed with mass $m_2$ of WD from $0.02M_\odot$ to $0.5M_\odot$, and the solid line shows the critical frequency $f_c$ changed with the mass of WD and the MT rate $\chi$ from $-2.2\times 10^{-7}M_\odot/\text{year}$ to $-2.2\times 10^{-5}M_\odot/\text{year}$. The mass of the primary star is $m_1=2M_\odot$.}
    \label{fRWD}
\end{figure}

To use the low-frequency approximation to solve Eq.\eqref{GWf} to obtain the frequency evolution, we must first determine the range of applicability of this method. As mentioned earlier, we can determine the radius of the white dwarf as well as the Roche limit from the mass of the WD, and then find the GW frequency $f_{RLOF}$ in the Roche limit
\begin{equation}
    f_{RLOF}=\frac{(GM)^{1/2}}{\pi} \left[\frac{0.6q^{2/3}+\ln(1+q^{1/3})}{0.49q^{2/3}}R_{WD}\right]^{3/2},
\end{equation}
where $R_{WD}$ is the radius of WD, we can use the Chandrasekhar radius-mass relation \cite{1972ApJ175417N} to calculate the radius 
\begin{equation}
\begin{split}
     R_{WD}\simeq&{4724~\text{km}}\left(\frac{M_{WD}}{0.7M_\odot}\right)^{-1/3}\\
    &\times \left[1-\left(\frac{M_{WD}}{M_{CH}}\right)^{4/3}\right]^{1/2}\left(\frac{\mu_e}{2}\right)^{-5/3},   
\end{split}
\end{equation}
where $M_{CH}=1.44M_\odot$ is the Chandrasekhar limit and $\mu_e=2$ is the mean molecular weight per electron. To use the low-frequency approximation to study the MT effect, the GW frequency should be lower than the critical frequency  $f_c$ given by Eq.\eqref{fc} and higher than the Roche frequency $f_{RLOF}$. The result is shown as the gray area in Fig.\ref{fRWD}. It can be seen that in the 10 mHz band, with the MT rate of $2.2\times 10^{-6} M_\odot$/year, the maximum WD mass is about 0.2$M_\odot$. 

Therefore, our low-frequency solution in the latter part of the paper does not have a strict demarcation of ‘low-frequency’, because the critical frequency $f_c$ depends on the mass of the binary, the rate of MT, and our low-frequency solution still applies to the frequency band up to 1Hz when the rate of MT reaches $1 M_\odot$/year.

\section{GW signatures with MT}\label{Sec3}
In the low-frequency case, the GW frequency is below the critical frequency Eq.\eqref{fc}, the MT effect dominates the frequency evolution, and the behavior of the time evolution of the GW frequency will be different from the situation without MT. Therefore, we focus on the low-frequency case in the remainder of this paper.

\subsection{Perturbation solution in the low-frequency case}
Eq.(\ref{ff}) in the low-frequency case when the rate of MT satisfies the condition 
\begin{equation}\label{lowf}
    \mathcal{F}\left(m_{1,0}-\chi t\right)\left(m_{2,0}+\chi t\right)f_0^{8/3}\ll\frac{3\mathcal{C}|\chi|}{m_{2,0}+\chi t},
\end{equation}
where $f_0$ is the initial frequency. Under these conditions, we can set the perturbation expression of GW frequency as
\begin{equation}
    f_L=f_{L0}+ f_{L1},
\end{equation}
and $f_{L0}\gg f_{L1}$. The subscript $L$  indicates low frequency, where $f_{L0}$ represents the dominant component of frequency $f_L$ due to MT, and $f_{L1}$ denotes the perturbative component induced by GWs. Substituting the above equation into Eq.\eqref{ff}, we obtain the leading order equation 
\begin{equation}
    \dot{f}_{L0}=3\mathcal{C}\chi(m_{2,0}+\chi t)^{-1}f_{L0},
\end{equation}
the solution of this  equation is
\begin{equation}
    f_{L0}=f_0\left(1+\frac{\chi t}{m_{2,0}}\right)^{3\mathcal{C}},
\end{equation}
where $f_0=f_{L0}(t=0)$. From Eq.(\ref{ff}), the equation for the next-to-leading order is given by
\begin{equation}
\begin{split}
    \dot{f}_{L1}
    =& \mathcal{F}\left(m_{1,0}-\chi t\right)\left(m_{2,0}+\chi t\right){f}_{L0}^{11/3}\\
    &+3\mathcal{C}\chi(m_{2,0}+\chi t)^{-1}f_{L1}.
\end{split}
\end{equation}
Note that $\mathcal{C}(q)$ is approximate to be constant, where $q=m_{2,0}/m_{1,0}$. Then we get the total expression of $f_L$ in low-frequency case
\begin{widetext}
\begin{equation}
    \begin{split}
        f_L&=-\frac{\mathcal{F} f_0^{11/3}}{\chi (8\mathcal{C}+3)(4\mathcal{C}+1)} \left\{\left(1+\frac{\chi t}{m_{2,0}}\right)^{11\mathcal{C}}\bigg[2(4\mathcal{C}+1)\chi^3 t^3+ \Big((16\mathcal{C}+3)m_{2,0}-(8\mathcal{C}+3) m_{1,0}\Big)\chi^2 t^2\right.\\
        &\left.+ \Big(8\mathcal{C}m_{2,0}^2-2 (8\mathcal{C}+3) m_{1,0}m_{2,0}\Big)\chi  t-m_{2,0}^2 \Big(m_{2,0}+(8\mathcal{C}+3) m_{1,0}\Big)\bigg]+\left(1+\frac{\chi t}{m_{2,0}}\right)^{3\mathcal{C}}\bigg[1+(8\mathcal{C}+3) m_{1,0}  \bigg]m_{2,0}^2\right\}. 
    \end{split}
\end{equation}
\end{widetext}

Since we focus on GWs in the sensitive band of the space-based detectors, $f_L(t)$ will not change significantly during the observing  run.
Therefore, we perform a Taylor expansion of the result $f_L(t)$ and keep the terms up to the order $\mathcal{O}(t^4)$. These terms are precise enough to perform the Fisher analysis. Thus, the GW frequency in the low-frequency case is
\begin{equation}\label{fLt}
    f_L(t)=f_0+f_{1t} t+\frac{1}{2!}f_{2t}t^2+\frac{1}{3!}f_{3t}t^3+\frac{1}{4!}f_{4t}t^4+\mathcal{O}\left(t^5\right).
\end{equation}
where
\begin{widetext}
\begin{align}
    f_{1t}&=3f_0\left(\frac{\chi}{m_{2,0}}\right)\mathcal{C} +\mathcal{F} f_0^{11/3}  m_{1,0}m_{2,0},\\
    f_{2t}&=  3f_0 \left(\frac{ \chi }{m_{2,0}}\right)^2\mathcal{C} (3 \mathcal{C}-1)+ \mathcal{F} f_0^{11/3}  \left(\frac{\chi}{m_{2,0}}\right)\Big[(14 \mathcal{C}+1) m_{1,0}-m_{2,0}\Big]m_{2,0},\\
    f_{3t}&=3 f_0 \left(\frac{\chi}{m_{2,0}}\right)^3 \mathcal{C} \Big[9 (\mathcal{C}-1) \mathcal{C}+2\Big]+\mathcal{F} f_0^{11/3} \left(\frac{\chi}{m_{2,0}}\right)^2\Big[\mathcal{C} (163 \mathcal{C}+8) m_{1,0}-(25 \mathcal{C}+2) m_{2,0}\Big]m_{2,0},\\
    f_{4t}&=9 f_0\left(\frac{\chi}{m_{2,0}}\right)^4 \mathcal{C} (\mathcal{C}-1) (3 \mathcal{C}-2) (3 \mathcal{C}-1)-2 \mathcal{F} f_0^{11/3}  \left(\frac{\chi}{m_{2,0}}\right)^3 \Big[\big(\mathcal{C} (51-910 \mathcal{C})+1\big) m_{1,0}+3 (73 \mathcal{C}+5) m_{2,0}\Big]m_{2,0}.
\end{align}
\end{widetext}
The terms proportional to $\mathcal{F}$ originate from the gravitational radiation back-reaction. It can be seen that the first term of each order is caused only by MT and $f_{it~1st~term}\propto \mathcal{C} f_0 (\chi/m_{2,0})^i$, which means if $\mathcal{C}=0$ this effect will vanish. These terms correspond to the direct influence of MT. The second term is caused by both MT and GWs, $f_{it~2nd~term}\propto \mathcal{F}f_0^{11/3} (\chi /m_{2,0})^{i-1}$ except $f_{1t~2nd~term}$ is caused only by GWs. The second term does not go to zero when $\mathcal{C} = 0$ but retains some of the effect caused by the combination of GWs and MT, which is due to the change of the chirp mass $\mathcal{M}_c(t)=\frac{(m_1m_2)^{3/5}}{M^{1/5}}$ in Eq.\eqref{GWf} as a result of MT. These terms correspond to the indirect influence of MT. However, \cite{2024arXiv240216612L} studies only the indirect effect of MT  and  \cite{10.1093/mnras/stad2358} only the direct effect. 

According to the low-frequency condition \eqref{lowf}, the magnitude of $f_{it~2nd~term}$ is smaller than the magnitude of $f_{it~1st~term}$. Moreover, $\chi$ is negative, so the frequency $f_L(t)$ decreases with time. The evolution of frequency is qualitatively different from the situation without MT. The phase of GWs can be calculated by integrating of the angular frequency $2\pi f_L(t)$ as $\psi_L=2\pi\int f_L(t)\mathrm{d}t$, then we get
\begin{equation}
        \psi_L(t)=\psi_{L,0}+2\pi \left[f_0 t+\sum_{i=1}^4 \frac{1}{(i+1)!}f_{it}t^{i+1}+\mathcal{O}\left(t^6\right)\right],\label{psiL}
\end{equation}
The difference in phases between the solutions with MT and without MT is given by
\begin{equation}
    \begin{split}
        \Delta\psi_{MT}(t)&=\psi_L(t)-\psi_{L,0}+\frac{3f_0^{-5/3}}{5\mathcal{F}m_{1,0}m_{2,0}}\\
        &-\frac{3}{5\mathcal{F}m_{1,0}m_{2,0}}\left(f_0^{-8/3}-\frac{8}{3}\mathcal{F}m_{1,0}m_{2,0}t\right)^{5/8},\label{DeltapsiL}
    \end{split}
\end{equation}

\begin{figure}[ht!]
    \centering
    {\includegraphics[width=0.48\textwidth]{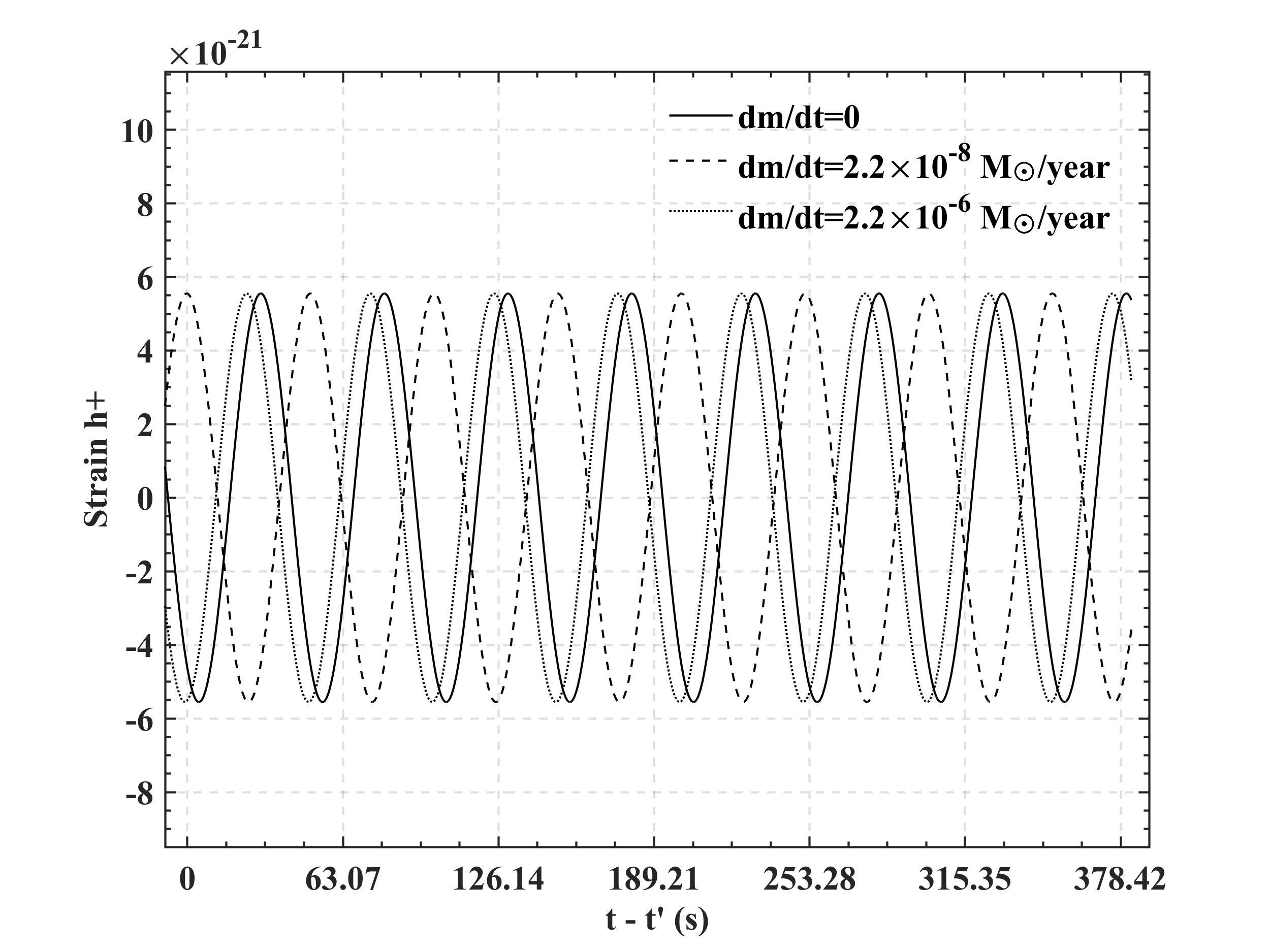}}
    \caption{Difference of phases of GWs with different initial MT rates, which is caused by cumulative effect of frequency change. The initial condition is $m_1=2.4M_\odot,m_2=0.025M_\odot,f=10$mHz and $\psi_{L,0}=0$. $t'$ equals two years, indicating that the differences in these waveforms are cumulative over two years. Note that the relative differences between the amplitudes are less than $10^{-2}$. The phase differences will accumulate to $\pi$ over two years of evolution.}
    \label{hh}
\end{figure}

Further, we can get the GW waveform when the angle between the line of sight and the angular momentum of the binary system is zero:
\begin{align}
    h_+(t)&=\frac{4}{\mathcal{D}}\left(\frac{G\mathcal{M}_c(t)}{c^2}\right)^{5/3}\left(\frac{\pi f_L(t)}{c}\right)^{2/3}\cos\psi_L(t),\label{ht}\\
    h_\times(t)&=\frac{4}{\mathcal{D}}\left(\frac{G\mathcal{M}_c(t)}{c^2}\right)^{5/3}\left(\frac{\pi f_L(t)}{c}\right)^{2/3}\sin\psi_L(t).\label{ht2}
\end{align}
where $\mathcal{M}_c(t)=(m_1m_2)^{3/5}/M^{1/5}$ is the chirp mass, $\mathcal{D}$ is the distance from the source to our detectors. The waveform is shown in Fig.\ref{hh}.  The relative difference in the amplitude between these three waveforms is less than $10^{-2}$.

It can be seen from the figure that when the accretion rate is the Eddington limit $\chi=-2.2\times 10^{-8}M_\odot\text{/year}$, the GW phase delays $\pi$ in two years, and when the accretion rate exceeds the Eddington limit $\chi=-2.2\times 10^{-6}M_\odot\text{/year}$, the GW phase delays more than $2\pi$. The detailed data will be shown in the next section.

\subsection{Analytical and numerical results}\label{ana_num}
To verify the correctness of the frequency function Eq.\eqref{fLt} in the low-frequency case we have calculated the error of $f_L$ and $\psi_L$ with $T=2$ years observation time, and results are shown as Table.\ref{f_L}
\begin{table*}[htbp!]
  \centering
  \caption{Comparison results of frequencies and phases. This table shows the initial frequency $f_0$, error between analytic and numerical frequency $\Delta f_L=f_L-f_{Num}$  and phase $\Delta \psi_L=\psi_L-\psi_{Num}$ and phase difference caused by MT $\Delta \psi_{MT}=\psi_L-\psi_{GW}$ with initial mass $m_{1,0}=2M_\odot$ frequency $f_0=1,10$ mHz.}
  \renewcommand\arraystretch{1.6}
    \setlength{\tabcolsep}{16pt}
    \begin{tabular}{c|c|cc|cc}
    \hline
    \hline
    &$f_0$ (mHz) & 1     & 10     & 1    & 10     \\
    \hline
    &$\chi (M_\odot/year)$  &  $-2.2\times10^{-6}$ &  $-2.2\times10^{-6}$  &  $-2.2\times10^{-7}$ &  $-2.2\times10^{-7}$   \\
    $m_2=$ &$\Delta f_L$(mHz)  & $-5.4\times10^{-11}$ & $-2.8\times10^{-8}$  & $-6.5\times10^{-12}$ & $-1.5\times10^{-8}$   \\
    $0.1M_\odot$ &$\Delta \psi_L$  & $6.1\times10^{-5}$ & $0.0045$  & $6.4\times10^{-6}$ & -0.0033  \\
    &$\Delta \psi_{MT}$  & -2.279 & -22.80  & -0.228 & -2.283   \\
        \hline
        \
    &$\chi (M_\odot/year)$   &  $-2.2\times10^{-6}$ &  $-2.2\times10^{-6}$  &  $-2.2\times10^{-7}$ &  $-2.2\times10^{-7}$   \\
    $m_2=$ &$\Delta f_L$(mHz)  & $2.1\times10^{-11}$ & $-9.4\times10^{-8}$  & $-4.4\times10^{-12}$ & $-8.4\times10^{-8}$   \\
    $0.25M_\odot$ &$\Delta \psi_L$  & $3.4\times10^{-5}$ & $0.0155$  & $1.6\times10^{-6}$ & 0.0144  \\
    &$\Delta \psi_{MT}$  & -1.343 & -13.44  & -0.1342 & -1.357   \\
    \hline
    \hline
    \end{tabular}%
  \label{f_L}%
\end{table*}%

\begin{figure*}[htbp!]
    \centering
    \subfigure{\includegraphics[width=0.48\textwidth]{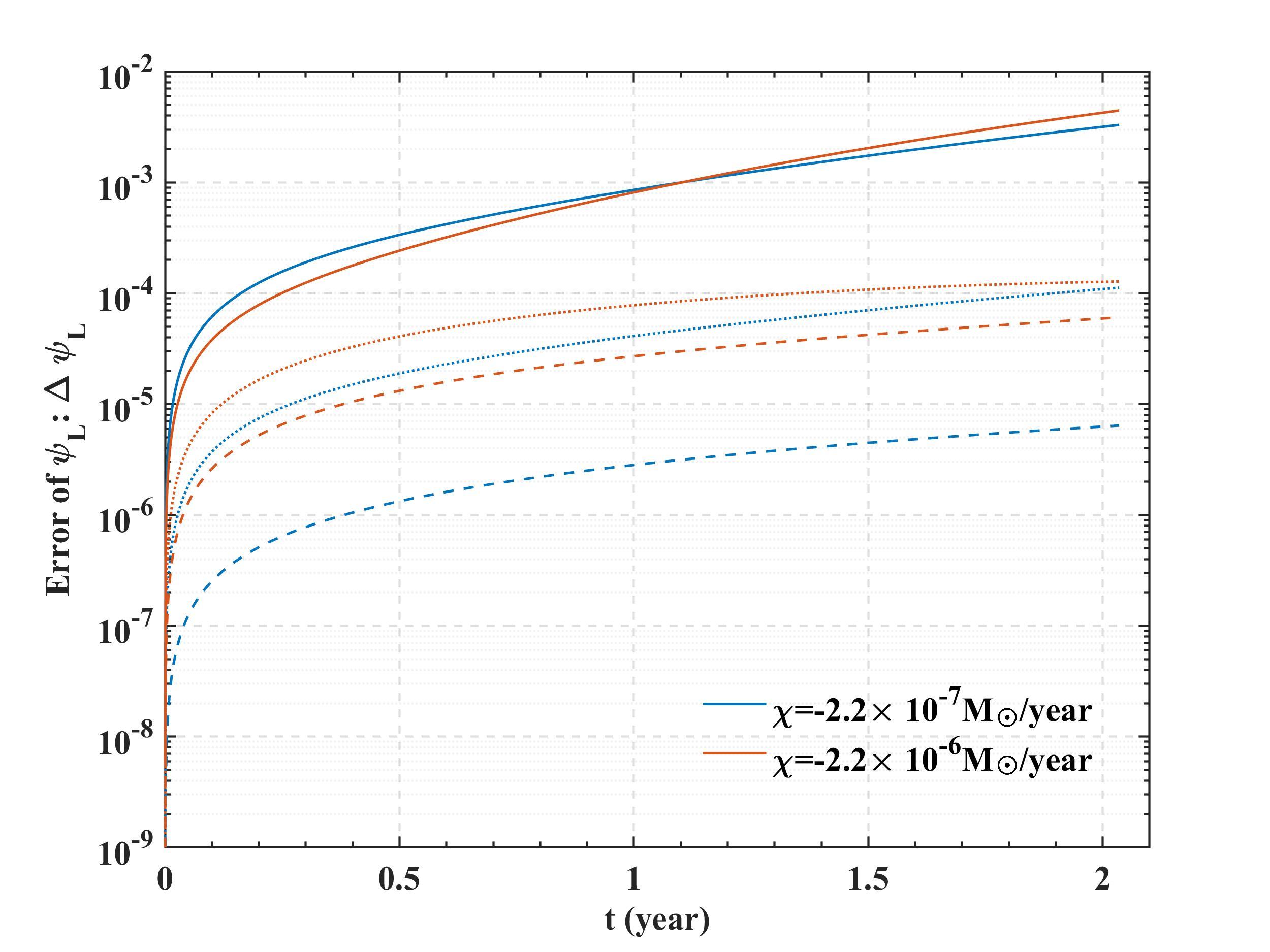}}
    \subfigure{\includegraphics[width=0.48\textwidth]{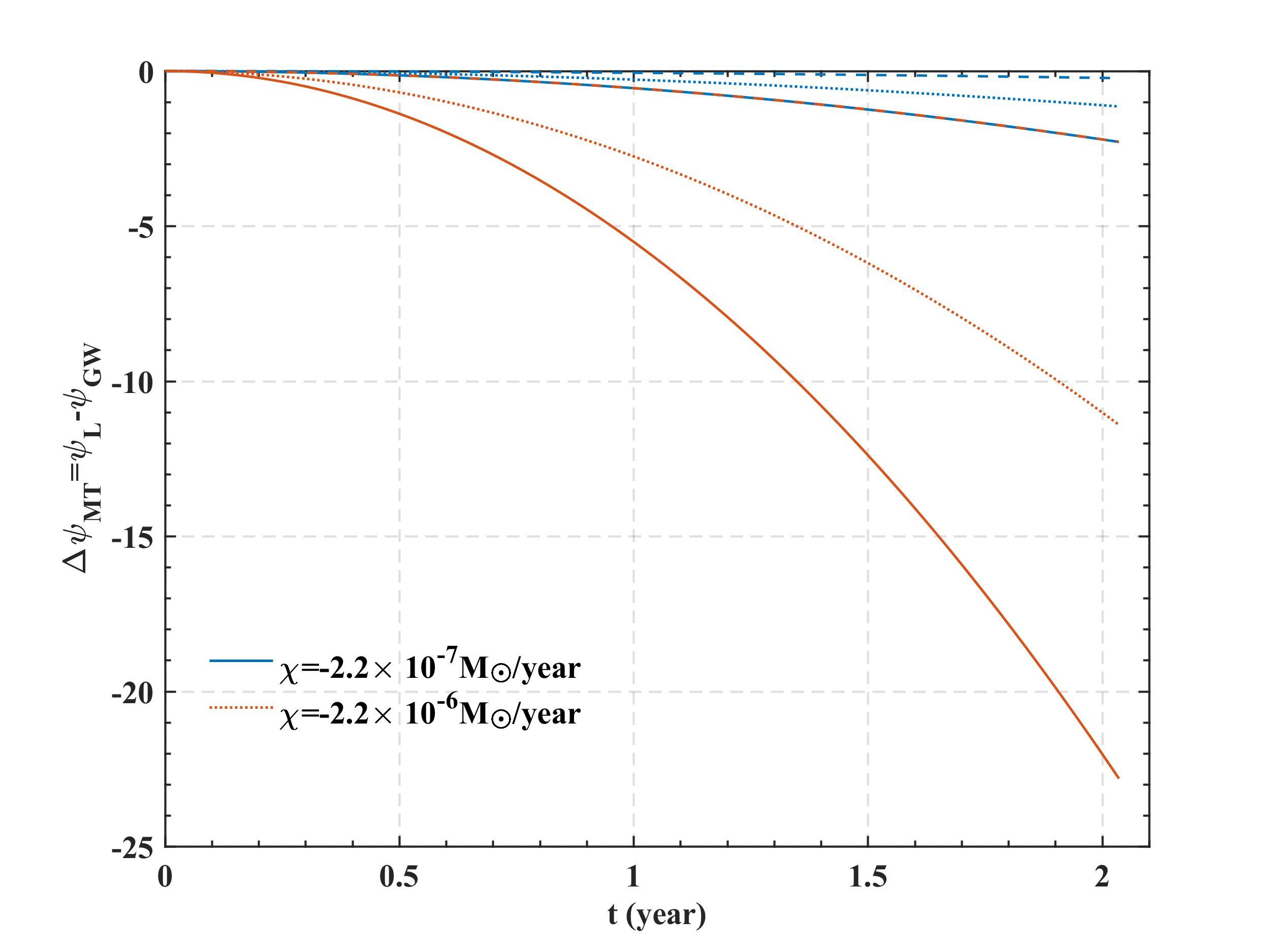}}\\
    \vspace{-0.2in}
    \subfigure{\includegraphics[width=0.48\textwidth]{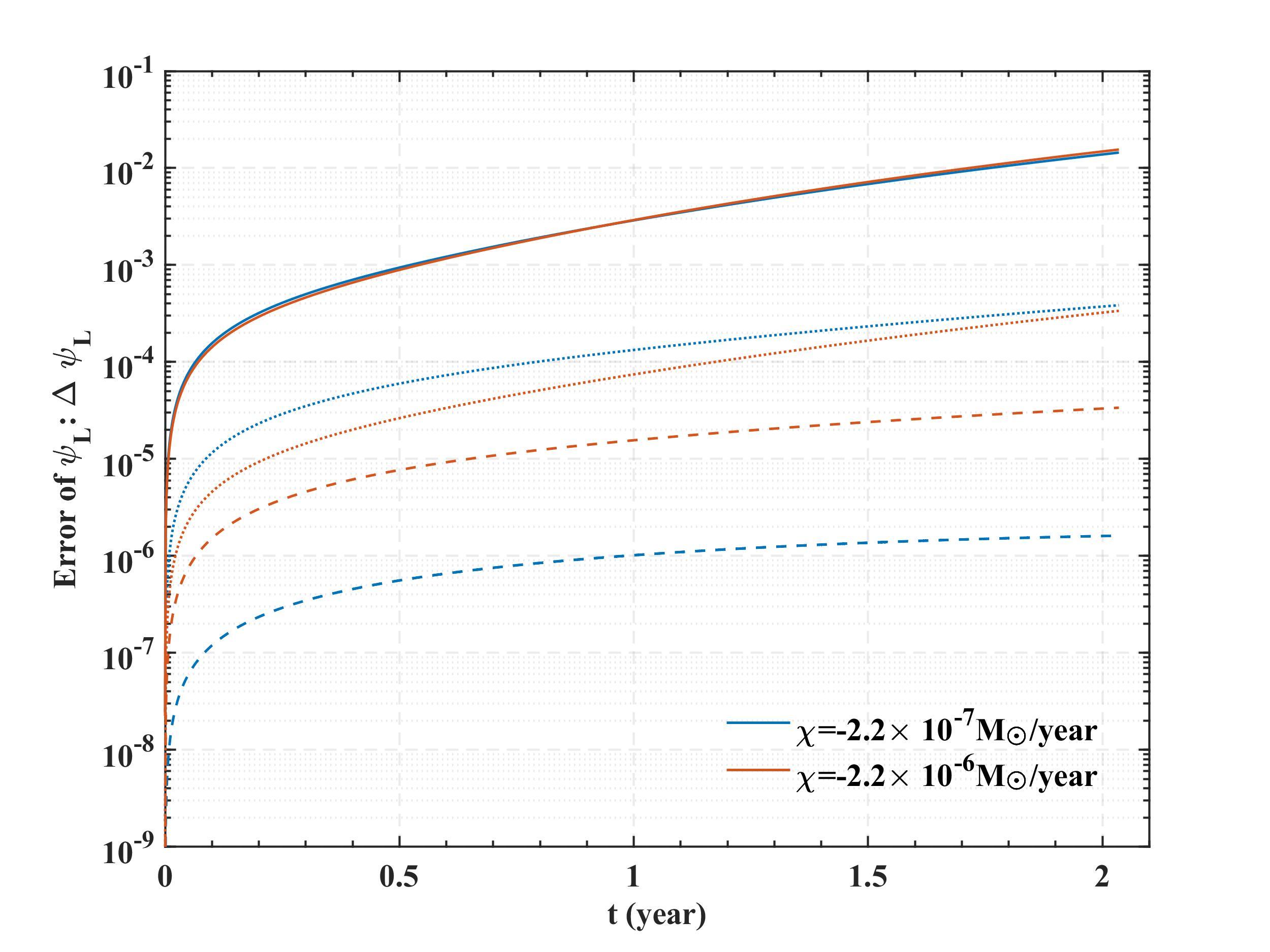}}
    \subfigure{\includegraphics[width=0.48\textwidth]{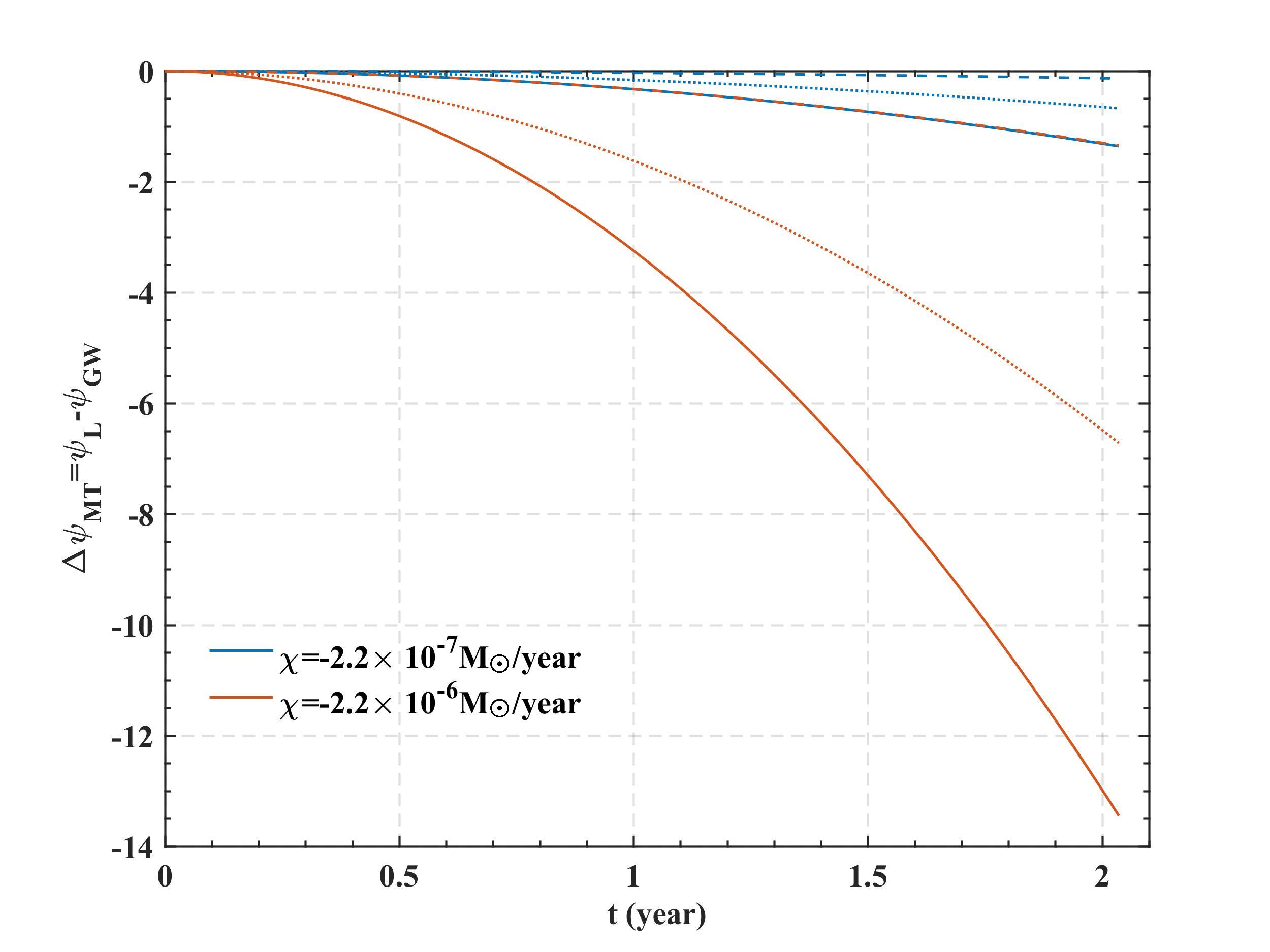}}
    \caption{Phase error between numerical results and analytic results (left). Phase difference with MT (right). $m_2=0.1M_\odot$ (up) and $m_2=0.25M_\odot$ (down). The dashed line corresponds to $f_0=1$ mHz, the dotted line corresponds to $f_0=5$ mHz, and the solid line corresponds to $f_0=10$ mHz. These figures show that when the MT rate $\chi=-2.2\times 10^{-7}M_\odot\text{/year}, -2.2\times 10^{-6}M_\odot\text{/year}$ the phase error caused by MT is much smaller than the phase difference caused by MT.}
    \label{error}
\end{figure*}

In Table \ref{f_L}, a preliminary estimate of the accuracy of the template can be made by determining how much the template changes to cause a $\pi$-radian change in the cumulative GW phase. We can see that MT brings a difference $\Delta \psi_{MT}>2\pi$ at the frequency  $10$mHz over 2 years of evolution. At the same time, the error between analytical and numerical results $\Delta \psi_L=\psi_L-\psi_{Num} $ is in the range $10^{-5}\thicksim10^{-2}$. Here, $\psi_{Num}$ is the phase calculated by numerically solving Eq.\eqref{GWf}. Thus, the phase \eqref{psiL} is accurate enough to quantify the MT effect. The difference in frequency is tiny which makes the strain of GWs change very small as well, but the phase change is larger because of the cumulative result of time. This table shows that the difference of phase will increase with the MT rate and this effect is more significant when WDs' mass is small.

More clearly, we plot $\Delta\psi_L$ and $\Delta\psi_{MT}$ in Fig.\ref{error}, which shows that the error is keeping small in the long time evolution and the difference of phase caused by MT is more and more noticeable. The MT rate of  blue lines is $\chi=-2.2\times 10^{-7}M_\odot\text{/year}$ and that of red lines is $\chi=-2.2\times 10^{-6}M_\odot\text{/year}$. They all have a similar trend, but the phase changes will increase larger when the MT rate increases. If the direct or indirect influence of MT is ignored, the accumulated GW phase difference can be greater than 1 radian.

\section{detectability of the MT effect}\label{Sec4}
\subsection{Response function of LISA and Taiji}
In this section we will use $h_+(t),\ h_\times (t)$ in Eq.(\ref{ht}) and Eq.(\ref{ht2}) to obtain the Michelson response $s(t)$ of the detector \cite{Gravity_PoissonWill,PhysRevD.67.022001}
\begin{equation}
s(t)=F_+(t)A_+(t)\cos\psi_L(t)+F_\times(t)A_\times(t)\sin\psi_L(t).
\end{equation}
where $ h_{+}(t)=A_+(t)\cos\psi_L(t), h_{\times}(t)=A_\times(t)\sin\psi_L(t)$. The response can be rewritten as
\begin{equation}
    s(t)=A(t)\cos[\psi_L(t)+\phi_D(t)+\phi_P(t)],\label{st}
\end{equation}
where $\Psi(t)=\psi_L(t)+\phi_D(t)+\phi_P(t)$, $A(t)=\sqrt{[A_+(t)F_+(t)]^2+[A_\times(t)F_\times(t)]^2}$, $\phi_D(t)$ is the frequency modulation and $\phi_P(t)$ is phase modulation.

The detector pattern functions $F_+(t),\ F_\times(t)$  are 
\begin{align}
    F_+(t)&=\frac{1}{2}\big[\cos2\zeta D_+(t)-\sin2\zeta D_\times(t)\big],\\
    F_\times(t)&=\frac{1}{2}\big[\sin2\zeta D_+(t)+\cos2\zeta D_\times(t)\big],
\end{align}
where $\zeta$ is the polarization angle of GWs, $D_+(t)$ and $D_\times(t)$ are function of the direction and position of the detector\cite{PhysRevD.57.7089}.

\subsection{Fisher metrix}
 To match the GW signals with the detection data, we need to have a realistic model of the detector noise $n(t)$. The remaining Gaussian noise can be described by its spectral density $S_n(f)$.

We can use a rough analytic fit to the noise curve of the detector as \cite{babak2021lisasensitivitysnrcalculations,Cao}
\begin{equation}
\begin{split}
    S_n(f)&=\frac{10}{3L}\left[1+\frac{3}{5}\left(\frac{f}{f_*}\right)^2\right]\\
&\times\Bigg[P_{OMS}+2\left(1+\cos^2\left(\frac{f}{f_*}\right)\right)\frac{P_{acc}}{(2\pi f)^4}\Bigg].
\end{split}
\end{equation}
For LISA \cite{Robson_2019} we have: 
\begin{equation}
    \begin{gathered}
        P_{OMS}=(1.5\times 10^{-11}\text{m})^2 \text{Hz}^{-1},\\ 
        P_{acc}=(3\times 10^{-15}\text{ms}^{-2})^2[1+(4\times 10^{-4}\text{Hz}/f)^2]\text{Hz}^{-1},\\
        L=2.5\times 10^9 \text{m},~ f_*=c/(2\pi L).
    \end{gathered}\nonumber
\end{equation} 
For Taiji \cite{PhysRevD.102.024089}: 
\begin{equation}
    \begin{gathered}
        P_{OMS}=(8\times 10^{-12}\text{m})^2 \text{Hz}^{-1},\\
        P_{acc}=(3\times 10^{-15}\text{ms}^{-2})^2[1+(4\times 10^{-4}\text{Hz}/f)^2]\text{Hz}^{-1},\\
        L=3\times 10^9 \text{m},~ f_*=c/(2\pi L).
    \end{gathered}\nonumber
\end{equation} 


The inner product $(s_1|s_2)$ of two signals $s_1(t)$ and $s_2(t)$ is \cite{PhysRevD.49.2658}
\begin{equation}
\begin{split}
   (s_1|s_2)&=2\int^\infty_{0}\frac{\Tilde{s}^*_1(f)\Tilde{s}_2(f)+\Tilde{s}_1(f)\Tilde{s}^*_2(f)}{S_n(f)}\mathrm{d}f\\
   &=\frac{2}{S_n(f_0)}\int^{T_{obs}}_{0}s_1(t)s_2(t)\mathrm{d}t, 
\end{split}
\end{equation}
where $\Tilde{s}(f)$ is the Fourier transforms of $s(t)$ and $T_{obs}$ is the observation time. 

When the  signal noise ratio (SNR) is large enough, the parameter-estimation errors $\Delta \theta^i$ can be determined by the Fisher matrix \cite{PhysRevD.46.5236}
\begin{equation}
    \Gamma_{ij}\equiv\left(\frac{\partial s}{\partial \theta^i}\bigg|\frac{\partial s}{\partial \theta^j}\right).
\end{equation}
The root-mean-square error in $\theta^i$ is
\begin{equation}
\sqrt{\left<(\Delta\theta^i)^2\right>}=\sqrt{\Sigma^{ii}},\;\;\;\bm{\Sigma}=\bm{\Gamma}^{-1}.
\end{equation}

In Eq.(\ref{st}), we consider only  6 independent parameters $\theta_i=\{\mathcal{D},\psi_{L,0},m_{1,0},m_{2,0},\chi,f_0\}$. Note that we set $\mathcal{C}(q) = 1-q$ and $r_h=0$ in the calculation of the Fisher matrix to reduce the possible error introduced by the numerically fitted accretion disk parameter $r_h$. The SNR is given by \cite{PhysRevD.49.2658,Takahashi_2002}
\begin{equation}
\begin{split}
    \left(\frac{S}{N}\right)^2&=4\int^\infty_0\frac{|\Tilde{s}(f)|^2}{S_n(f)}\mathrm{d}f\approx\frac{2}{S_n(f_0)}\int^{T_{obs}}_0 |s(t)|^2\mathrm{d}t.
\end{split}
\end{equation}
In the following, we choose the observation distance $\mathcal{D}=1$kpc for parameter estimation and the corresponding SNR is in the range $10^2\sim 10^3$. 

\subsection{Numerical results}
\begin{figure}[ht!]
    \centering
    {\includegraphics[width=0.48\textwidth]{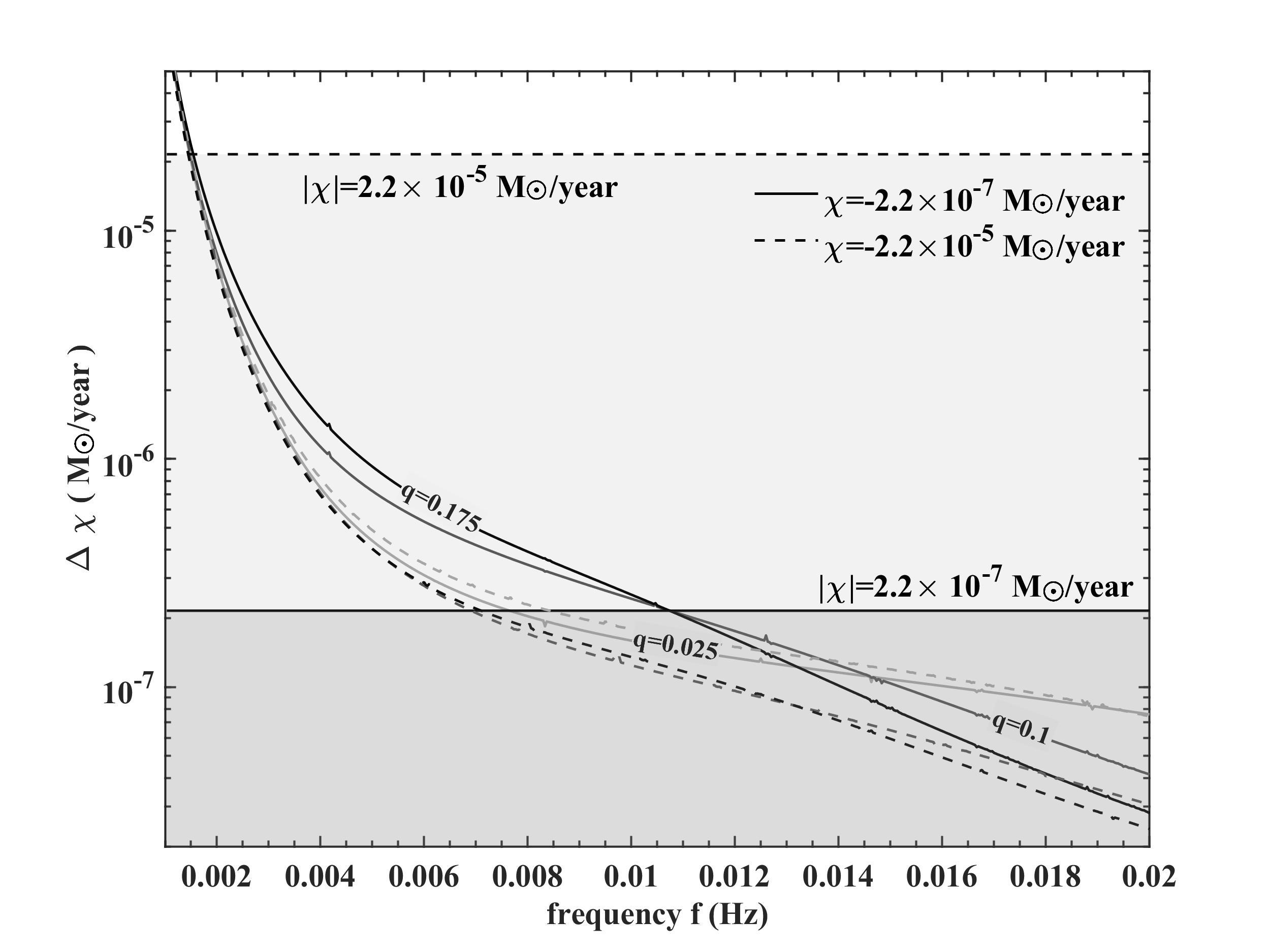}}
    \caption{Uncertainties of the MT rates of different detection scenarios with  Taiji. The distance is $\mathcal{D}=1$ kpc. The mass of the primary star is $m_1=2M_\odot$.  The mass ratios are $q=0.025,0.1,0.175$. The MT rates are $\chi=-2.2\times10^{-5}M_\odot/\text{year}$ ( dotted lines ) and $\chi=-2.2\times10^{-7}M_\odot/\text{year}$ ( solid lines ). When the solid line enters the dark gray area below or the dotted line enters the light gray area below, $\Delta \chi<|\chi|$ and the MT effect can be detected. }
    \label{MT_f_Chi}
\end{figure}

We calculate  the Fisher matrix $\bm{\Gamma}$ and get its inverse matrix and the parameters' error $\Delta \theta_i$. We consider 6 parameters $\{\mathcal{D},f_0,\psi_{L,0},m_{1,0},q,\chi\}$\cite{Cao,ebadi2024lisadoublewhitedwarf}. Part of the result is in Table.\ref{fisher}. The measurement uncertainties $\Delta \chi$ in measuring the MT rate $\chi$ for a binary at 1 kpc between $2\sim 20$ mHz by Taiji are shown in Fig.\ref{MT_f_Chi}.
 
It can be seen from Fig.\ref{MT_f_Chi} that when the MT rate is the Eddington accretion limit of NS $\chi=-2.2\times10^{-7} M_\odot/$year (solid lines) and the frequency  $f_L \gtrsim 10$ mHz, $\Delta \chi<|\chi|$ and the MT effect can be detected by the GW observation. This means that the MT effect is detectable in the NS-WD binary within 1 kpc with the GW frequency higher than 10 mHz. 

\begin{table*}[htbp!]
\small
  \centering
  \caption{SNR and the uncertainties of parameter estimation for  binaries with MT of different detection scenarios with LISA and Taiji}
    \renewcommand\arraystretch{1.5}
    \setlength{\tabcolsep}{11pt}
    \begin{tabular}{c|ccc|ccc}
    \hline
    \hline
          &  $\theta_{i,0}(1)$ &  LISA &  Taiji &  $  \theta_{i,0}(2)$ &  LISA &  Taiji \\
    \hline
    $\mathcal{D}~(\text{kpc})$     & $1$ & $0.0060$ & $0.0031$  & $1$ &$0.0037$ & $0.0017$ \\
    $f_0~$(Hz) & $\bm{5\times10^{-3}}$ & $4.8\times10^{-11}$ & $2.5\times10^{-11}$ & $\bm{1\times10^{-2}}$ & $3.8\times10^{-11}$&$1.7\times10^{-11}$ \\
     $\psi_{L,0}$ & 0     & 0.0082 & 0.0042     & 0 & 0.0053     & 0.0024 \\
    $m_{1}~(M_\odot)$ & $\bm{2}$  & $2.0571$ & $1.0672$     & $\bm{2}$ & $0.1513$     & $0.0700$ \\
    $q$ & 0.1   & 0.1767 & 0.0917   & 0.1 & $0.0130$   & $0.0060$ \\
     $\chi~(M_\odot/\text{year})$ & $\bm{-2.2\times 10^{-7}}\!\!$& $1.4\times 10^{-6}$ & $7.2\times 10^{-7}\!\!$ & $\bm{-2.2\times 10^{-7}}\!\!$ & $5.3\times 10^{-7}\!\!$ & $2.4\times 10^{-7}$ \\
    SNR& &287.3 & 553.8 & &462.4 & 999.2\\
    \hline
     &  $\theta_{i,0}(3)$ &  LISA &  Taiji &  $  \theta_{i,0}(4)$ &  LISA &  Taiji \\
    \hline
    $\mathcal{D}~(\text{kpc})$    & $1$ & $0.0139$ & $0.0064$ & $1$ & $0.0143$ & $0.0066$  \\
    $f_0~$(Hz) & $\bm{1\times10^{-2}}$ & $2.3\times10^{-10}$ & $1.1\times10^{-10}$& $\bm{1\times10^{-2}}$ & $2.7\times10^{-10}$ & $1.3\times10^{-10}$ \\
     $\psi_{L,0}$ & 0     & 0.0102 & 0.0047 & 0     & 0.0103 & 0.0048  \\
    $m_{1}~(M_\odot)$ & $\bm{1}$   & $1.4479$ & $0.6707$  & $\bm{1}$   & $0.8430$ & $0.3901$ \\
    $q$ & 0.2   & 0.5101 & $0.2363$ & 0.2   & 0.2963 & $0.1371$ \\
     $\chi~(M_\odot/\text{year})$ & $\bm{-2.2\times 10^{-7}}\!\!$& $7.0\times 10^{-6}$ & $3.2\times 10^{-6}$ & $\bm{-2.2\times 10^{-6}}\!\!$& $6.3\times 10^{-6}$ & $2.9\times 10^{-6}$ \\   
    SNR& &283.0 & 661.4 & &283.0 & 661.4\\
    \hline
    \hline
    \end{tabular}%
  \label{fisher}%
\end{table*}%

Comparing $\theta_{i,0}(1)$ and $\theta_{i,0}(2)$ in Table.\ref{fisher}, it can be seen that all the uncertainties decrease when the initial frequency $f_0$ increases from 5 mHz to 10 mHz.
$\theta_{i,0}(3)$ and $\theta_{i,0}(4)$ are the parameters of binary WDs. Comparing these two parts, it can be seen that the uncertainties of the mass $m_1$ and the mass ratio $q$ decrease by about $40\%$ when the MT rate increases by a factor of 10. With the increasement of the MT rate, $\Delta m_1< m_1$ and $\Delta q < q$ for the detection by Taiji and the individual masses of binary WDs can be measured. Without the MT effect, GWs emitted by binary WDs are almost monochromatic. This makes it challenging to measure the individual masses of WDs \cite{10.1093/mnrasl/slaa183}. Our result shows that when the MT effect is incorporated, the individual masses can be measured.
 \begin{figure}[ht!]
    \centering
    {\includegraphics[width=0.48\textwidth]{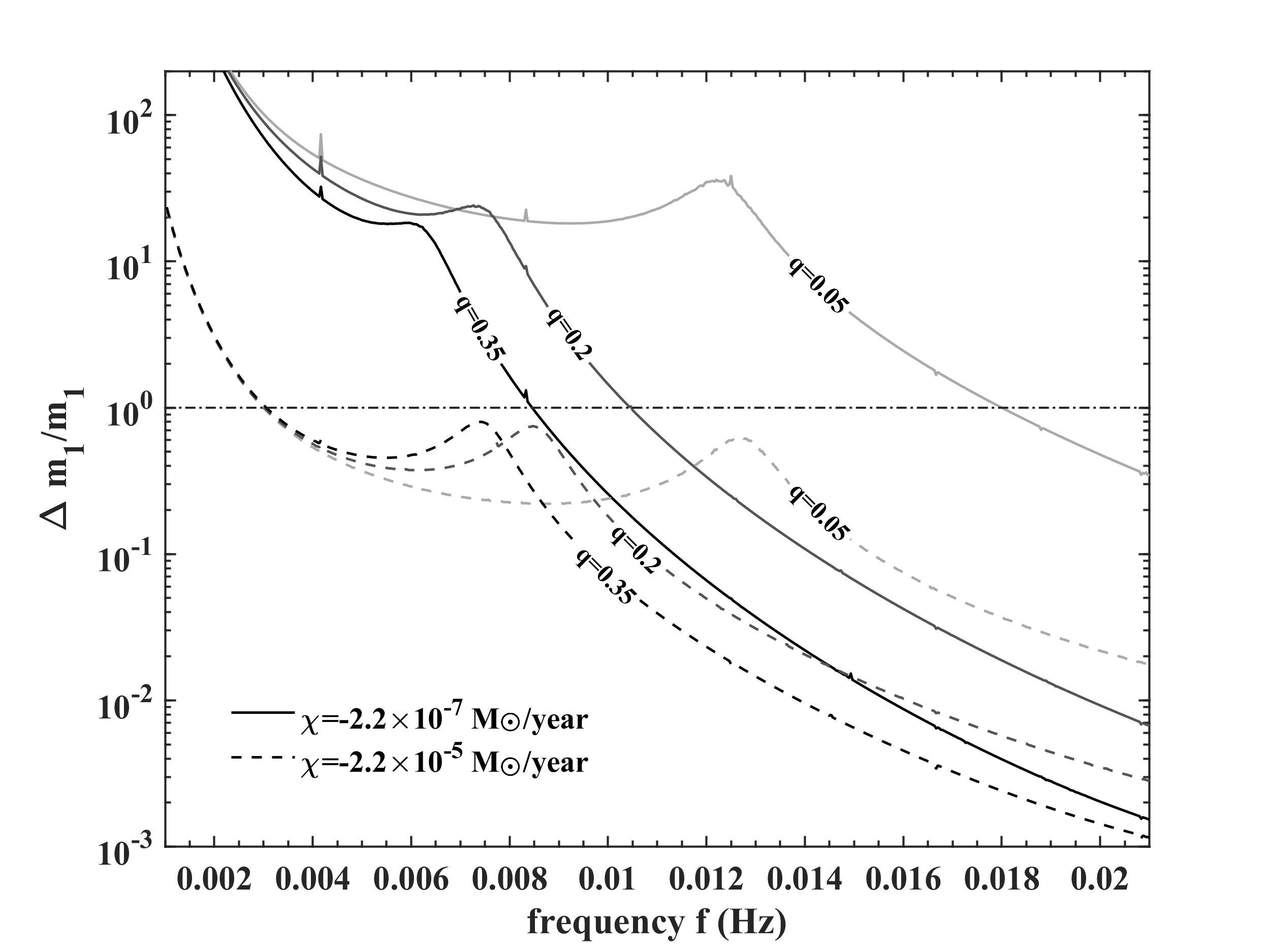}}
    \caption{Relative uncertainties of the mass of the primary star in double WDs of different detection scenarios with Taiji. The mass of the primary star is $m_1 = 1 M_\odot$. The mass ratios are $q = 0.05, 0.2, 0.35$, and the MT rates are $\chi=-2.2\times 10^{-5} M_\odot/$year and $-2.2\times 10^{-7} M_\odot/\text{year}$. If the relative uncertainty is less than $\Delta m_1/m_1=1$, the mass of the primary star can be measured.}
    \label{Deltam}
\end{figure}


To show more clearly the help that the MT effect brings to the measurement of the binary mass,  we plot in Fig.\ref{Deltam} the relative uncertainties of the mass of the
primary star $\Delta m_1/m_1$ in double WDs detected by Taiji. As can be seen from the figure, $m_1$ can be measured in the 5 mHz band when the MT rate reaches $|\chi|\thicksim 10^{-5}M_\odot/\text{year}$, and it can also be measured in the 10 mHz band when the accretion rate is $|\chi|\thicksim 10^{-7}M_\odot/\text{year}$. In addition, for a given mass of the primary star, a larger mass of the companion star will make the measurement easier.

\section{Conclusions and discussions}\label{sec5}
We have considered the MT effect in the gravitational waveforms and studied the detectability of this effect. Assuming that the MT effect on the orbital evolution is dominant over the gravitational radiation, we obtain the analytical expression for the GW frequency and waveforms. When the MT rate is below the Eddington limit, the impact of this effect on the gravitational waveforms is mainly in the phase. We adopt Fisher analysis to compute the projected uncertainties of the MT rate. We find
that for NS-WD binaries within the distance of 1 kpc in frequency 10 mHz, when the MT rate reaches $2.2\times10^{-7}M_\odot/\text{year}$, the MT effect can be detected by Taiji. For double WDs within the distance of 1 kpc in frequency 10 mHz, if the MT rate is high enough ($ \gtrsim 2.2\times10^{-7}M_\odot/\text{year}$), the individual masses can be measured with Taiji.

In the previous section, when calculating the Fisher
matrices, the angles in the detector pattern functions are
not incorporated. To study the influence of these angles,
we have generated 100 samples of the GW sources by
randomly generating these angles while keeping  the other
parameters the same as $\theta_{i,0}(2)$ in Table \ref{fisher}. Then, incorporate these angles into the
Fisher matrices. We find that both the average of the
uncertainties of the mass $m_1$ and the mass transfer rate
increase by about $13\%$. Due to the degeneracy between
the angles and the source distance, the average uncertainties of the source distance becomes 13 times larger
than in the case where no angles are considered. That
is to say, incorporating these angles does not change the
conclusions about the measurement of the mass transfer
rate and the mass.

It can be seen from Fig. \ref{Deltam} that when the GW frequency is high enough, the relative mass uncertainties can be less than 0.1, and the conclusion of this paper is based on the high-frequency part of Fig. \ref{Deltam}. In addition, the SNR of the GW sources of this part is always greater than 100, so that the Fisher matrix approximation is valid in this situation.

The Fisher matrix approximation is still valid when the
mass uncertainties are comparable to the masses themselves. We have changed the value of the mass transfer
rate from $2.2\times10^{-5}M_\odot/\text{year}$ to $2.2\times10^{-7}M_\odot/\text{year}$ and
kept other parameters fixed ($\mathcal{D}=1\text{kpc}$, $\psi_{L,0}=0$, $f_0=0.005\text{Hz}$, $m_1=2 M_\odot$, and $q=0.2$). As the mass transfer rate
decreases, the relative uncertainties of the mass increase
from $3\%$ to $103\%$. However, the uncertainties of the distance, phase and frequency change by less than $1\%$. This
means that the Fisher matrix approximation is still valid
to obtain these uncertainties when the mass uncertainties
become comparable to the masses themselves.

Various extensions exist for future work. We have assumed the MT rate to be constant. It would be interesting to study the influence of the variation of the MT rate on the waveforms \cite{10.1093/mnras/stab626}. It would also be important to incorporate the tidal interaction into the waveforms \cite{10.1093/mnrasl/slaa183}. Furthermore, one may extend the current analysis by considering the
orbital eccentricity and higher post-Newtonian correction to study the 
orbital dynamics in more detail \cite{Gravity_PoissonWill}, since we only consider the circular orbit to the Newtonian order.

\section{ACKNOWLEDGMENTS}
T. L. is supported by the China Postdoctoral Science Foundation Grant No. 2024M760692. This work is supported in part by the National Key Research and Development Program of China Grant No. 2020YFC2201501, No. 2020YFC2201404, and No. 2021YFC2203003, in part by the National Natural Science Foundation of China under Grant No. 12075297 and No. 12235019.
\bibliography{BIB.bib}
\bibliographystyle{unsrt}

\end{document}